\documentclass{epl}
\usepackage{amsmath}
\usepackage{graphics}
\def\bchi{\bm{\chi}}
\def\Tr{\mathrm{Tr}}
\title{Modified TAP equations  for the SK spin glass}
\author{T. Plefka}
\institute{
 Theoretische Festk\"orperphysik, TU Darmstadt, D
64289 Darmstadt, Germany} 
\begin{document}
\maketitle
\begin{abstract}
The stability of the TAP mean field equations is  reanalyzed with
the conclusion that the exclusive reason for the breakdown at the
spin glass instability is an inconsistency for the value of the
local susceptibility. A new alternative  approach leads to
modified equations which are in complete agreement with the
original ones above the instability. Essentially altered results
below the instability are presented and the consequences for the
dynamical mean field equations are discussed.
\end{abstract}
\section{Introduction}
Together with the replica approach, the Thouless-Anderson-Palmer
(TAP) approach \cite{tap} is the most important method  to analyze
infinite range spin glass models like the Sherrington-Kirkpatrick
(SK) model \cite{sk} of Ising spins (for reviews see
\cite{mpv,fh}). The TAP equations are well established
\cite{mpv,fh,I} and are exact in the thermodynamic limit ($N
\rightarrow \infty $) provided that the local magnetizations
$\{m_i\}$ satisfy the condition
\begin{equation}\label{5}
 x\equiv 1-\beta^2( 1-2q_2+q_4)>0\qquad \textrm{where}\qquad
 q_\nu=N^{-1}\sum_i m_i^\nu \qquad \nu=2,4.
\end{equation}
The condition (\ref{5}) represents the central stability condition
for the SK spin glass and therefore is also found in other
approaches \cite{at,mpv,fh}. Within the TAP approach two different
arguments for (\ref{5}) are known.  Bray and Moore \cite{bm} found
a divergence of the spinglass susceptibility for $ x\rightarrow0
$. The expansion of Plefka \cite{I}, leading to the TAP equations,
is limited to $ x>0$.

In next section a further, basically simple, aspect of the
stability condition is presented, leading naturally to a
modification of the TAP equations.
\section{Stability analysis revised}
The TAP free energy $F$ of the SK model is given by
\begin{eqnarray}\label{1}
 F(\beta,\{m_i\})=  -\frac{1}{2} \sum_{ij} J_{ij}m_i m_j -\frac{N\beta}{4}(1-q_2)^2 -\sum_i h_im_i\nonumber\\
+ \frac{1}{2\beta}\sum_i \Big\{(1+m_i)\ln \frac{1+m_i}{2}+(1-m_i)
\ln \frac{1-m_i}{2}\Big\}.
\end{eqnarray}
where the $\{h_i\}$ are local external fields.  The bonds $
J_{ij}$ are independent random variables with zero means and
standard deviations $ N^{-1/2}$ . From $\partial F/\partial m_i=0$
the TAP equations
\begin{equation}\label{3}
m_i=\tanh \beta\Big\{h_i +\sum_j J_{ij}m_j-\beta m_i(1-q_2)\Big\}
\end{equation}
result and the stability is governed by  the inverse
susceptibility matrix
\begin{eqnarray}\label{4}
 \chi_{ij}^{-1}=\partial h_i/\partial m_j=\partial^2 F/\partial
m_i \partial m_j =\delta_{ij} \Big\{\beta^{-1}(1-m_i^2)^{-1} +
\beta( 1-q_2)\Big\} -J_{ij}
\end{eqnarray}
where the $N^{-1}$- order term $- 2N^{-1}\beta m_i m_j $ has been
dropped as it has no influence on the present analysis (compare,
however, \cite{I} and the footnote below).

According to (\ref{1}), the free energy is  formally well defined
for all values of the $\{m_i\}$ inside the hyper cube $ |m_i|\leq
1$ including the regime $ x<0 $. The question is  which physical
requirement is violated for  $ x<0 $. To answer this question we
investigate the free energy for convexity, a fundamental
requirement for every free energy function. This can be done
without explicit knowledge of the solutions of (\ref{3}). The
analysis is performed  analogously   to \cite{I} and the resolvent
of the matrix $\bchi^{-1}$ is introduced
\begin{equation}\label{6}
R(z)\equiv N^{-1}\,\mathrm{Tr} \,(z-\bchi^{-1})^{-1}.
\end{equation}
It is important to note (compare appendix) that $ R(z) $ is an
element of the class of functions which are analytic in $ z$ for $
\mathrm{Im}z\neq0 $ and such that
\begin{equation}\label{x}
  \mathrm{Im}\;R(z)>0 \qquad \mathrm{for}\qquad\mathrm{Im}\,z<0 \:.
\end{equation}

The key for  further  analysis is the powerful theorem of Pastur
\cite{pastur}, rederived  in \cite{bm} as ' locator expansion' .
According to (\ref{a1}) the resolvent $R(z)$ satisfies the
equation
\begin{equation}\label{7}
R(z)=N^{-1}\sum_i \big\{z-R(z)-\beta^{-1}(1-m_i^2)^{-1} - \beta(
1-q_2)\big\}^{-1}
\end{equation}
in the  $N \rightarrow \infty $ limit for nearly all
configurations  of the bonds (which does not imply  averaging).
Eq. (\ref{7}) is basically a polynomial of the order $ N+1 $ and
has therefore $ N+1 $ solutions for $R(z)$ . With reference to the
appendix and to \cite{pastur}, however, there is only one solution
which satisfies (\ref{x}). Thus $R(z)$ can uniquely be determined.

Near $ x=0$ and $ z=0 $, the leading behavior of $R(z)$  is
obtained by a power expansion of the right hand site of (\ref{7})
in terms of $ \{R-z +\beta( 1-q_2)\} $ . To second order
\begin{equation}\label{8}
 p\,\big\{R-z +\beta( 1-q_2)\big\}^2 \,+ \,x \,\big\{R-z +\beta( 1-q_2)\big\} \,+\,z \,=\,0
\end{equation}
is found, where $ p=\beta^3 N ^{-1}\sum_i  (1-m_i^2)^3 $ is a
positive quantity. For the case $p|z|\ll x^2$, the solutions
$R^\pm(z)$ are given by
\begin{equation}\label{9}
 \{R^+-z +\beta( 1-q_2)\}=-z/x \quad \textrm{and by}\quad
\{R^--z +\beta( 1-q_2)\}= -x/p+ z/x
\end{equation}
where according to the  requirement (\ref{x}) \textit{the
solution}
 $R^+(z)$ \textit{applies for} $x>0$ and \textit{ the solution} $R^-(z)$ \textit{applies for} $x<0$,
respectively.

Several important conclusions can be made. According to (\ref{6})
\begin{equation}\label{10}
  \chi_l\equiv- \textrm{Re}\,R(z)\Big |_{z\rightarrow 0}= N^{-1}\,\mathrm{Tr} \, \bchi
\end{equation}
holds for the local susceptibility $ \chi_l$ , which gives, with
eq. (\ref{9})
\begin{equation}\label{11}
 \chi_l=\beta( 1-q_2) \quad \textrm{for } x>0
\qquad\textrm{and}\qquad \chi_l=\beta( 1-q_2)+x/p \quad
\textrm{for}\; x<0\;,
\end{equation}
respectively . \textit{The value of} $ \chi_l$ \textit{deviates
from the value} $\beta( 1-q_2)$ for $ x<0$. The latter value,
however, is exact for an arbitrary  Ising model in the canonical
distribution as can easily  be shown. Moreover, in general, the
exact value is  used at the beginning of the derivations leading
to the TAP equations. Thus  \textit{the inconsistent result $
\chi_l\neq \beta( 1-q_2)$ causes the breakdown of the TAP approach
for $ x<0$}.

The spin glass susceptibility $\chi_{sg}$ is, according to
(\ref{6}), related to $ R(z) $ by
\begin{equation}\label{12}
 \chi_{sg}\equiv- \textrm{Re}\,\frac{\partial R(z)}{\partial z}\Big |_{z\rightarrow 0}=
N^{-1}\,\mathrm{Tr} \, \bchi^2
\end{equation}
from which in leading order
\begin{equation}\label{13}
\chi_{sg}= |x|^{-1}  \qquad  (x\neq 0)
\end{equation}
is obtained. Apart from the divergence, $\chi_{sg}$ is well
behaved and  positive everywhere. The earlier work \cite{bm,fh} is
therefore restricted to the case $ x>0$, although this is not
explicitly stated. Within the replica method $\chi_{sg}$ becomes
negative for $ x<0$ \cite{mpv,fh}. According to (\ref{9}), a
negative $\chi_{sg}$ only results if (\ref{x}) is violated.

The eigenvalue density $ \varrho(\lambda) $ of $ \bchi^{-1}$ is
determined from $R(z)$ by
\begin{equation}\label{14}
\varrho(\lambda)= \pi^{-1}\,\mathrm{Im}\; R(\lambda-i
\epsilon)\big|_{\epsilon\rightarrow +0}\quad .
\end{equation}
With the full solution of the quadratic eq.(\ref{8}) one obtains
for small values of $\lambda$ and of $ x $
\begin{equation}\label{15}
\varrho(\lambda)= \pi^{-1}p^{-1/2}\sqrt{\lambda-x^2/p}.
\end{equation}
This result shows that the minimum eigenvalue of $ \bchi^{-1}$ is
given to leading order by
\begin{equation}\label{16}
  \lambda_{min}=\,x^2/p \, \geq0 \qquad \mathrm{for}\quad  |x|\rightarrow 0 .
\end{equation}
With the extension to arbitrary values of $x$  given in the
appendix, the last result implies that negative eigenvalues do not
occur and the \textit{TAP free energy is semi-convex} everywhere.
Note that the present analysis and thus the results are limited to
the leading order in $N$. Hence the TAP free energy may not be
semi-convex everywhere on sub-extensive scales. Such an interesting
behavior was found for the p-spin spherical model \cite{cgp} and
is expected in regions near $ x=0$ due to finite size effects of
$\varrho(\lambda)$. \footnote{In this context the sub-extensive
terms to $\bchi^{-1}$ neglected in eq. (\ref{4}) and  leading to a
second stability condition $1 -2 \beta^2 ( q_2-q_4)>0 $ (compare
\cite{I}) may have some effect despite the conclusions of \cite{owen}
that these terms are not important.}
\section{The modified TAP equations} It was shown that the TAP
equations break down for $ x<0 $ due to an inconsistency for the
value of $ \chi_l$. We therefore look for modified equations which
do not show such an unsatisfactory behavior. The present analysis
starts with the well founded expressions \cite{tap,mpv,fh} which
are rederived   in the appendix
\begin{equation}\label{17}
 m_i=\tanh \beta\big\{h_i +\sum_j J_{ij}m_j-m_i \chi_l\big\}
\end{equation}
where the local susceptibility is given by $ \chi_l=N^{-1}\sum_i
\chi_{ii}= N^{-1}\sum_i
\partial{m_i}/\partial{h_i}$. With the working hypothesis that $
\partial{\chi_l}/\partial{m_i} $ is a order $ N^{-1} $ term, the
inverse susceptibility matrix $ \chi_{ij}^{-1}=
\partial{h_i}/\partial{m_j}$ is  calculated from eqs.(\ref{17}) to
\begin{equation}\label{19}
\chi_{ij}^{-1}=\delta_{ij} \{\beta^{-1}(1-m_i^2)^{-1} + \chi_l\}
-J_{ij}.
\end{equation}
Note that \textit{the eqs.}(\ref{17}),(\ref{19}) \textit{and} $
\sum_j \chi_{ij}\chi_{jk}^{-1}=\delta_{ik}$ \textit{represent a
closed set of self consistent equations} for  the $ m_i$, the
$\chi_{ij} $ and the $\chi_{ij}^{-1}$. This fact is the key to the
present approach. Moreover it is important to realize that
\textit{already the} eqs.(\ref{17}) \textit{are complete and
sufficient to determine all solutions} as the eqs.(\ref{19})
result from eqs.(\ref{17}). In this context the restriction of the
original TAP equations to $x>0$ is obvious as the replacement
$\chi_l\rightarrow \beta(1-q_2)$ is only justified for stable
solutions but will lead to inconsistencies for other solutions.

The theorem of Pastur leads to a very  effective simplification of
the above set of equations. Application of this theorem yields $
R(-i\epsilon)=-N^{-1}\sum_i \big\{\beta^{-1}(1-m_i^2)^{-1}+
 R(-i\epsilon) +\chi_l +i \epsilon\big\}^{-1}$ where  $ R(z) $
 is now related to the  $\bchi^{-1}$ given in
 eq.(\ref{19}). Performing  the limit $\;\epsilon\rightarrow +0$, taking again
care of the  requirement (\ref{x}), noting that $\chi_l=
-\textrm{Re}\,R(-i \epsilon) |_{\epsilon\rightarrow +0}$ holds and
separating  the real  and imaginary parts finally results in
\begin{eqnarray}\label{22}
\chi_l=\beta(1-q_2)\qquad,\qquad \Gamma=\,0\;&  &\mathrm{for}\,x\geq 0\\
\chi_l =\frac{1}{N}\sum_i
\frac{\beta(1-m_i^2)}{1+\Gamma^2\,\beta^{2}(1-m_i^2)^2}\qquad,\qquad
 1 =\frac{1}{N}&\sum_i
\frac{\beta^2(1-m_i^2)^2}{1+\Gamma^2\,\beta^{2}(1-m_i^2)^2}\qquad
 &\mathrm{for}\,x\leq 0\label{23}
\end{eqnarray}
where $ \Gamma= \pi^{-1}\rho(0)=\textrm{Im}\,R(-i \epsilon)
|_{\epsilon\rightarrow +0}\geq 0$ was introduced. Note that both
$\chi_l $ and $\Gamma$ are continuous at $x$ and note the latter
eq. of (\ref{23}) has always a solution $\Gamma$ for $x<0$.
Finally it is easy to show that the above working hypothesis is
satisfied.

The set of eqs.(\ref{17}), (\ref{22}) and (\ref{23}) represent the
modified TAP equations which determine $ \{m_i\}, \chi_l $ and $
\Gamma $ for all temperatures and all magnetic fields. In the
region $ x>0$, the modified and the original TAP equations and
thus their solutions are identical. Dramatic changes, however, are
found for the region $ x<0 $ in which the equations differ,
leading in general to different solutions. The stability of the
solutions is governed by   the value of $ \Gamma $  which is
proportional to the eigenvalue density $ \rho(0)$ . In the region
$ x>0$ all solutions are stable with $ \Gamma=0$ and in the regime
$ x<0 $ all solutions are unstable with $ \Gamma>0$.

A solution of special interest is the paramagnetic solution $
m_i=0$ in zero field. This solution satisfies the modified TAP
equations with $\Gamma=(1-T^2)^{1/2}
>0$  for $ T<1$  and with $ \Gamma=0$ for $ T>1$, respectively. Thus
the solution is unstable for temperatures below $T=1$. This
physically important result is  not found by the original TAP
approach which gives a stable paramagnetic solution for all
temperatures.

Eq.(\ref{22}) shows that for all stable solutions  the value of $
\chi_l$  equals  the thermodynamic value $ \beta(1-q_2) $. A
difference  of these two values occurs for the unstable solutions.
This is not in conflict with thermodynamics which is a priori
limited to stable states.

The last argument, the general limitation of equilibrium
thermodynamics to stable states, implies that for unstable states
thermodynamic functions like the entropy or the free energy are
not defined \footnote{Discussions based on free energy landscapes
use often \textit{extensions of   thermodynamic free energies } to
the instable regimes.  Thus such approaches are phenomenological.
For the present work such an obvious extension is not available
and the basically equivalent approach via relaxational equations
of motions is employed.} . Again from the strict thermodynamic
point of view the original and the modified TAP equations are
equivalent as both lead to the same results for the stable states.
It is  the reason for the breakdown which differs in both
approaches. The inconsistency for $\chi_l$ for $x<0$ of the
original TAP equations  is rather unconventional compared to the
modified equations where the usual behavior, negative eigenvalues
of $\bchi^{-1}$, applies.
\begin{figure}
\onefigure[height=5cm]{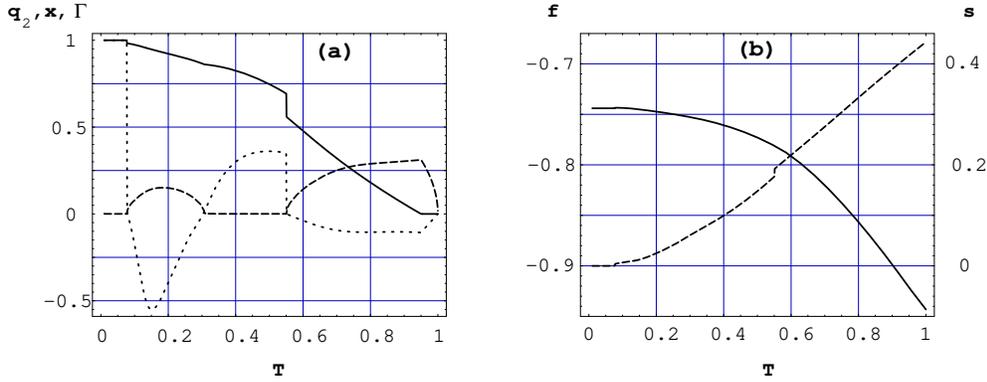}
\caption{\textbf{Zero field
solution of the modified TAP equations. }
  \textbf{(a)} Edwards Anderson parameter $q_2 $ (full line),
  $ x\equiv 1-\beta^2( 1-2 q_2+q_4)$(dotted line), $ \Gamma$ (dashed line),
  \textbf{(b)} free energy density $f=F/N$ (full line) and entropy
  density $s$ (dashed line) versus temperature $ T $ for a system
  with $ N=100 $ and with $ J_{ij}=\pm N^{-1/2}$. The solution is
  obtained from the integration of (\ref{24})  by slowly cooling
  down from $T=1$. The discontinuities at $T=.078$ and $T=.551$
  result from saddle-node bifurcations with jumps to lower values of
  $ f$ (which are not resolved on the scale of (b)). Slow heating
  leads to hysteresis effects near the discontinuities.}
  \label{dat}
\end{figure}

Unstable states are important for phenomenological extensions to dynamics.
On basis of the modified TAP equations and in contrast to the usual
TAP treatment such a phenomenological but
consistent extension to the dynamics becomes possible.  In analogy to \cite{dyn,fh} the relaxational
Glauber dynamics in mean field approximation will be used for this
purpose. Measuring the time in units of the relaxation time, the
equations of motion are given by
\begin{equation}\label{24}
 \dot{m}_i(t) = - m_i(t)+ \tanh \beta \big \{h_i +\sum_j
J_{ij}m_j(t)-m_i(t) \chi_l(t) \big \}
\end{equation}
where $ \chi_l(t)$ again is determined by eqs.(\ref{22})
and (\ref{23}). The fixpoints of (\ref{24}) coincide  with the
solutions of static equations and a one-to-one correspondence
between the dynamical stability of the fixpoints and the
thermodynamic stability holds. Moreover the eqs.(\ref{24})
simulate the  evolution of thermodynamic processes in
which $\beta$ or $h_i$  are changed. Using the
initial magnetizations of such a process as initial values for the
time integration, the system   relaxes to a stable
fixpoint which corresponds to a stable solution of the static
TAP equations and  thus to the final state of the thermodynamic
process. The dynamical approach uniquely determines the final
state even in those cases where the system exhibits meta- or
multi-stability. From the pure knowledge of the static solutions
such a determination of the final state is not possible.

The final states are stable and thus they are always located
outside the regime $ x<0$. During the dynamic evolution, however,
the system may temporarily stay  or may even start in the region $
x(t) <0 $. Thus the flow of  eq.(\ref{24}) is needed in both
regions $ x(t)>0 $ and $ x(t) <0$. These arguments demonstrate the
relevance of the unstable states. Note that
equations of motion of the above type but  based on the original TAP equations usually
lead to incorrect results in the spinglass regime as the system
relaxes to paramagnetic solutions.

As the eqs.(\ref{24})  are phenomenological they do  not
describe the true dynamical effects of the SK model but can be used
as a numerical tool to find explicitly the solutions of the TAP equations for finite
$N$. Numerical work which is based on this
method and  which is an  alternative approach to \cite{nt},
will be published separately \cite{II}.

To demonstrate that this method is successful some results of
\cite{II} are presented in fig.(\ref{dat}). A system of $N=100$
spins  with $h_i=0$ and with binary distributions of the $J_{ij}$
was investigated. By numerical integration of eqs.(\ref{24}) with
(\ref{22}) and (\ref{23}) the fixpoints of these equations have
been calculated in the temperature range from $T=1$ down to $
T=.01$  with a step size of $\delta T=.01$.  The first run at
$T=1$  has performed with the initial values $ m_i=0$. All other
runs use as initial values the fixpoint values of the former run
to simulate a slowly cooling down. The results for $ q_2 , x,
\Gamma $,  the free energy and the entropy density are plotted in
fig.(\ref{dat}). Due to finite size effects \cite{bm,nt,II} the
boundary for the stability $x=0$ is not exact for finite $N$. Thus
as shown in  fig.(\ref{dat}) stable fixpoint solutions with $
\Gamma>0$ and with $ x<0$ are found. With increasing $ N$,
however, the boundary for stability tends to $x=0$  as
demonstrated in \cite{II}.
\section{Conclusions}
Based on a careful stability analysis it has been worked out that
the TAP equations need a modification in the region of
instability. This  modification leads to a complete and consistent
description of the spin glass instability including the instable
region. Although the new, unstable states cannot be observed, the
unstable regime essentially determines  the flow of the equation
of motion and thus affects the basins of attraction of the stable
states.

The semi-convexity of the TAP free energy  in the thermodynamic
limit represents a further result of importance as it implies that
the 'multi-valley' structure of the TAP free energy  occurs on a
sub-extensive scale. For further investigations of these important
effects the modified equations are expected to be an adequate
tool. It is straightforward to extend the analysis from the SK
model to the other numerous spin glass models of infinite range
and results similar to those of the present work are expected.
\section{Appendix:}
In the first point in this section  eq.(\ref{17}),
which is the starting point for the derivation to the modified TAP
equations is rederived. Similar to the approaches \cite{tap,mpv} the $N-1$ spin
system is considered, with the  Ising spin $ S_n $ removed from
the $ N $ spin system under investigation. Taking the trace over the
spin $S_n$ leads to the exact relations for all $ i(\neq n)$
\[m_i=\hat{m}_i + \frac{1+m_n}{2\beta}\frac{\partial a_n^+}{\partial
h_i} + \frac{1+m_n}{2\beta}\frac{\partial a_n^-}{
\partial h_i}\quad, \quad
m_n= \frac{\exp{( a_n^++ \beta h_n)}-\exp{(  a_n^- - \beta h_n)}}{
\exp{( a_n^++ \beta h_n)}+\exp{(  a_n^-- \beta h_n)}}
\]
where $\hat{m}_i= \langle S_i\rangle_{N-1} $ and
$\exp(a_n^\pm)= \langle\exp(\pm\beta \sum J_{ni}S_i)\rangle_{N-1}$
are expectation values of the $N-1$ spin system. Note that latter
of these values can be rewritten as a ratio of two partition
functions with fields $ h_i\pm J_{ni}$ and with fields $h_i$,
respectively. With the remark that the next steps are only
justified to the leading order in $ N^{-1}$ one finds by a
cumulant expansion $ a_n^\pm = \pm \beta \sum_i J_{ni}\hat{m}_i
+\beta \sum_{ij}J_{ni}J_{nj}
\partial{\hat{m}_i}/\partial{h_j} $. This leads to eq.(\ref{17})
with
$\chi_l=\sum_{ij}J_{ni}J_{nj}\partial{\hat{m}_i}/\partial{h_j}$
which can to leading order be approximated to $ \chi_l=N^{-1}
\sum_i
\partial{m_i}/\partial{h_i}$ due to the random character
of the $J_{in}$.

The theorem of Pastur \cite{pastur} is central for the present
work and in the remaining part of this appendix two technical
aspects related to this theorem are worked out. Let $ K_{ij}= k_i
\delta_{ij}$ be a non-random matrix in  $N$ dimensional space with
all $ k_i$ real valued  and let $ J_{ij}$ be a symmetric matrix
(with $ J_{ii} =0$), where the off-diagonal elements are
independent random quantities with zero means and standard
deviations $ N^{-1/2}$. According to this theorem the resolvent $
R(z)= N^{-1}\Tr\{z-\mathbf{K}+ \mathbf{J}\} ^{-1} $ satisfies the
equation in the limit $ N\rightarrow \infty $
\begin{equation}\label{a1}
 R(z)= N^{-1}\Tr
\big\{z-\mathbf{K}-R(z)\big\}^{-1}=N^{-1}\sum_i
\big\{z-k_i-R(z)\big\}^{-1}.
\end{equation}
\begin{figure}
\onefigure[height=3.5cm]{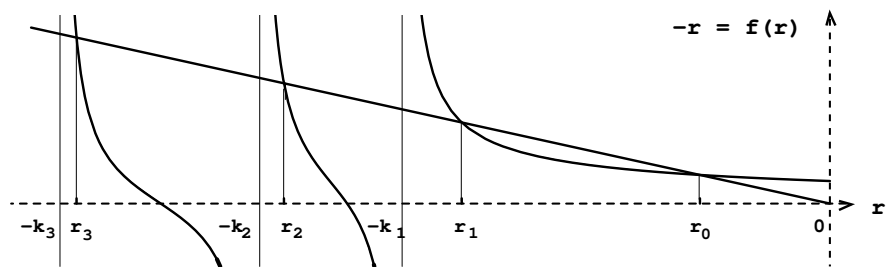} \caption{ Graphical  solution
of the equation $-r= N^{-1}\sum_i(r+k_i)^{-1}\equiv f(r)$. The
solutions $r_0$ and $r_1$ may be absent depending on the values of
the $k_i$.}
\label{pastur}
\end{figure}

Focusing on the properties of  the $ N+1$ solutions of (\ref{a1})
  $f(x)=N^{-1}\sum_i \big\{k_i+x\big\}^{-1} $ is introduced and $
k_1<k_2\ldots<k_N$ is presumed. Setting $ r=R(z=0)$, the solutions
of (\ref{a1}) for the case $z=0$ are determined by $ -r=f(r)$.
According to fig.(\ref{pastur}) there are always $ N-1 $ real
solutions $r_i \;(<-k_1)$ for $ i=2,\ldots N $ . Depending on the
values of $ k_i $ two cases are possible for the two remaining
solutions: For case (i) both solutions are real and will be
denoted by $ r_1 $ and $ r_0 $ (with $ r_1<r_0$). For case (ii) a
pair of conjugate complex solutions $ c^\pm $ results with Im
$c_+>0$ and with Im $c_-<0$, respectively.

In the next step the linear terms in $z$  of the solutions  $
R_i(z) $ of (\ref{a1}) are calculated by expansion for $ |z|\ll1$
near the real solutions $ r_i$. The ansatz $ R_i(z)= r_i +a_i \,z
$ leads in linear approximation to $-r_i -a_i\, z = f(r_i +a_i\,
z- z)\approx f(r_i)+ f'(r_i)(a_i-1)\,z$ from which
$a_i=f'(r_i)\{1+f'(r_i)\}^{-1}$ results. Fig.(\ref{pastur}) shows,
that $-1<f'(r_0) <0 $   and $ f'(r_{i\neq0}) <-1 $ holds and in
consequence   $ a_0<0 $ and $ a_{i\neq0}> 0$ results. Thus with
the secondary requirement Im $R(z)>0$  for Im $z <0$, the solution
of (\ref{a1}) is unique determined  and given in the limit $
z\rightarrow0 $ by $ r_0$ in case (i) and by $ c_+$ in case (ii),
respectively. The definition of $ R(z)$ satisfies this requirement
as $ \mathrm{Im}\,R(z)=-N^{-1}(\mathrm{Im}\,z)\sum_l
\big\{(\Lambda+\mathrm{Re}\,z )^2 +(\mathrm{Im}\,z)^2\big\}^{-1} $
holds, where $ \Lambda $ are the real eigenvalues of
$\mathbf{-K+J}$. Thus the solution of (\ref{a1}) together with the
requirement (\ref{x}) determines $ R(z) $ uniquely.

Let us finally consider the special points $\bar{\lambda}$  on the
real axis  in the $z$ plane where the real solutions $ R(
\bar{\lambda}) =\bar{r}$   of (\ref{a1}) bifurcate to  complex
values.  The values $\bar{\lambda}$ and $ \bar{r}$  are determined
by the solutions of $  \bar{r}=-f(\bar{r}-\bar{\lambda}) $ and of
$ f '(\bar{r}-\bar{\lambda})=-1 $. The first equation is just the
theorem (\ref{a1}) while the latter can be obtained by a
discussion similar to fig.(\ref{pastur}).  The  definition of
$f(x)$  leads to the identity $ f(u) -f(v) + (v
-u)f'(v)=N^{-1}(v-u)^2 \sum_i \;(k_i+u)^{-1}( k_i+v)^{-2}$ .
Focusing to the original TAP case
 $ k_i=  \beta^{-1}(1-m_i^2)^{-1} + \beta(1-q_2) $ has to
be used according to (\ref{7}) and (\ref{a1}). Setting  $ u= -
\beta(1-q_2) $ and setting  $ v=\bar{r}-\bar{\lambda}$ the
identity and the above  equations  lead to the important result
$\bar{\lambda}\geq0$ . This result implies that all endpoints of
the intervals with a finite $ \rho(\lambda)$ are located on the
positive axis. A $1/z$ -expansion of (\ref{6}) shows that $
\rho(\lambda)\equiv0$  for large negative $\lambda$ . Thus the
minimum eigenvalue satisfies $ \lambda_{min}\geq 0$ which
generalizes the result (\ref{16}) to all values of $x$.

\end{document}